\documentclass[twocolumn,amssymb, nobibnotes, aps, prd,amsmath,mathtools, nofootinbib,pgfmath,pgffor,superscriptaddress]{revtex4-1}
\usepackage{graphicx}

\newcommand{\be}{\begin{equation}}
\newcommand{\ee}{\end{equation}}
\makeatletter
\def\qrr@split@result#1 #2\@qrr@split@result{\edef\erfInput{#1}\edef\erfResult{#2}}
\newcommand*{\gnuplotErf}[2][\jobname.eval]{%
    \immediate\write18{gnuplot -e "set print '#1'; print #2, erf(#2);"}%
    \everyeof{\noexpand}
    \edef\qrr@temp{\input{#1} }%
    \expandafter\qrr@split@result\qrr@temp\@qrr@split@result
}
\makeatother

\usepackage{caption,color}
\usepackage{subcaption}
\usepackage{multirow}

\usepackage{hyperref}
\hypersetup{colorlinks,citecolor=blue}

\usepackage[dvipsnames]{xcolor}
\usepackage{color, colortbl}

\begin{document}

\title[Echoes from the Abyss: A status update]{Echoes from the Abyss: A Status Update}

\author{Jahed Abedi}
\email{jahed.abedi@aei.mpg.de}
\affiliation{Max-Planck-Institut f\"ur Gravitationsphysik, D-30167 Hannover, Germany}
\affiliation{Leibniz Universit\"at Hannover, D-30167 Hannover, Germany}

\author{Niayesh Afshordi}
\email[]{nafshordi@pitp.ca }
\affiliation{Department of Physics and Astronomy, University of Waterloo,
200 University Ave W, N2L 3G1, Waterloo, Canada
}
\affiliation{Waterloo Centre for Astrophysics, University of Waterloo, Waterloo, ON, N2L 3G1, Canada}
\affiliation{Perimeter Institute For Theoretical Physics, 31 Caroline St N, Waterloo, Canada}


\begin{abstract}
Gravitational wave echoes provide our most direct and surprising observational window into quantum nature of black holes.  Three years ago, the first search for echoes from Planck-scale modifications of general relativity near black hole event horizons led to tentative evidence at false detection probability of 1\% \cite{Abedi:2016hgu}. The study introduced a naive phenomenological model and used the public data release by the Advanced LIGO gravitational wave observatory for the first observing run O1 (GW150914, GW151226, and LVT151012, now GW151012). Here, we provide  a status update on various observational searches for echoes by independent groups, and argue that they can all be consistent if echoes are most prominent at lower frequencies and/or in binary mergers of more extreme mass ratio.  We also point out that the only reported ``detection'' of echoes (with $>4\sigma$ confidence) at 1.0 second after the binary neutron star merger GW170817 \cite{Abedi:2018npz} is coincident with the formation time of the black hole inferred from electromagnetic observations.  
\end{abstract}

\maketitle

\section{Introduction}
\begin{figure}
\begin{center}
\includegraphics[width=0.4\textwidth]{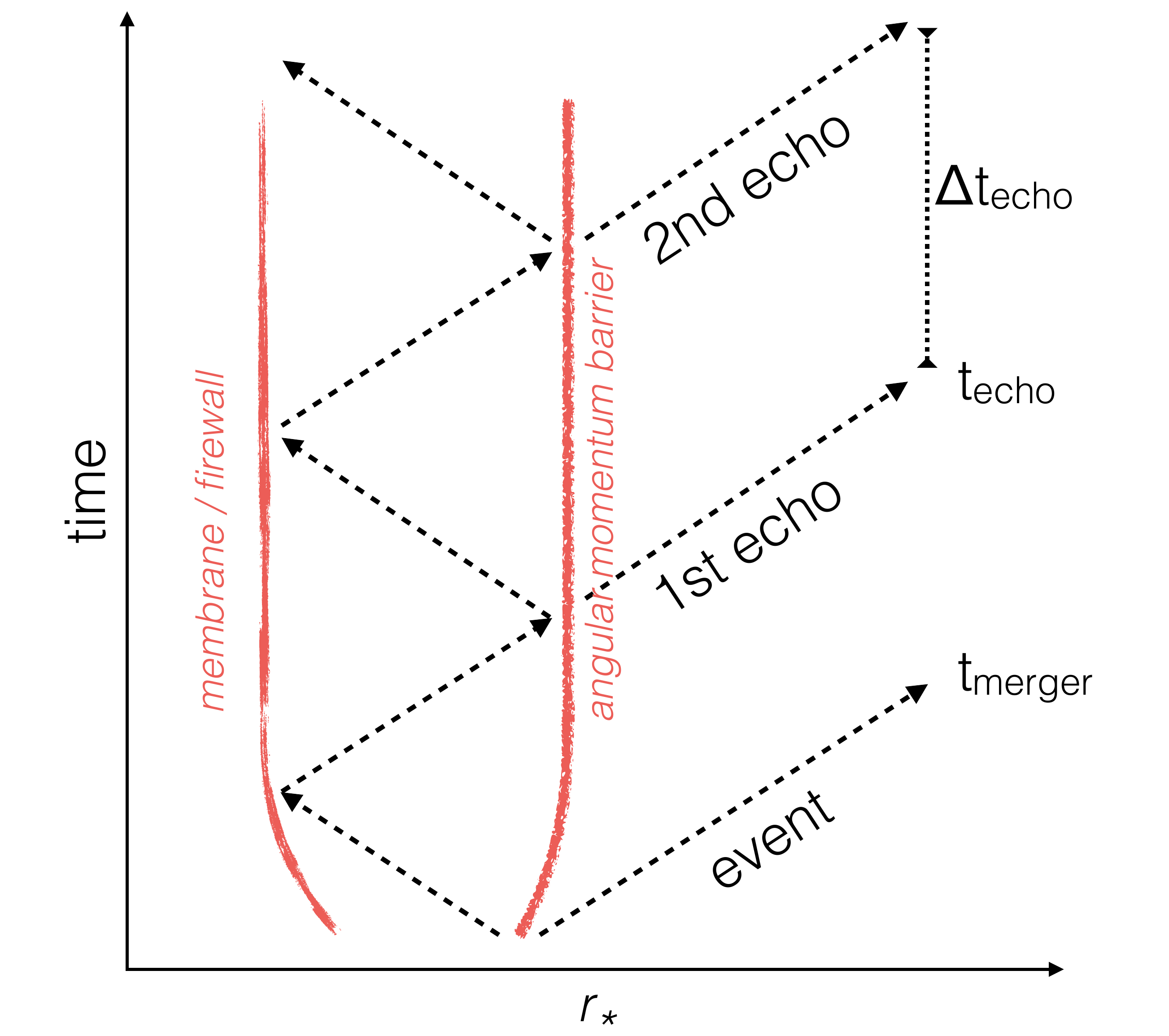}
\caption{Gravitational wave echoes following a BBH merger from a cavity of membrane/firewall-angular momentum barrier \cite{Abedi:2016hgu}. }
\label{echo_pic_1}
\end{center}
\end{figure}

\begin{figure}
\centering
\includegraphics[width=0.4\textwidth]{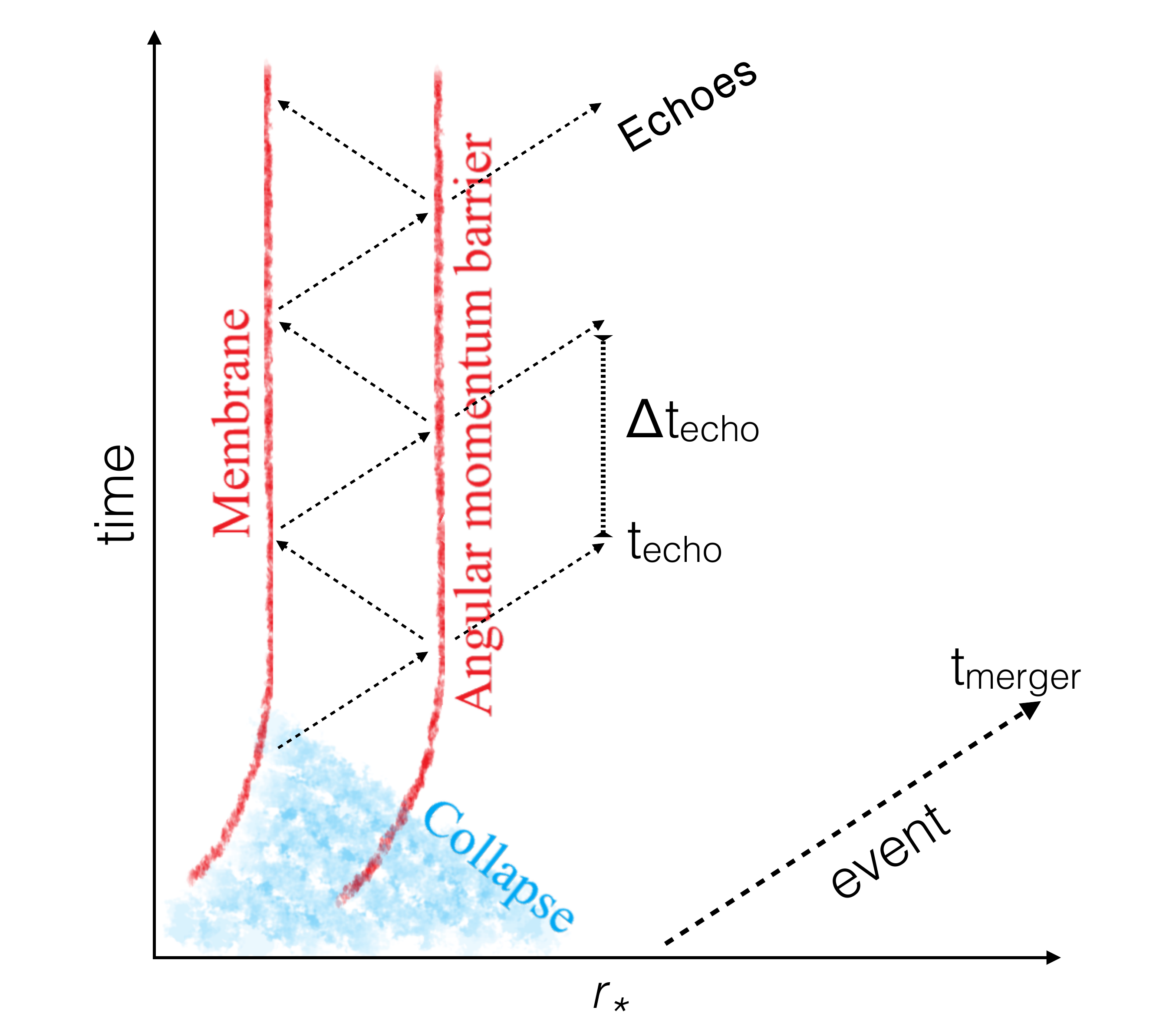}
\caption{Gravitational wave echoes following a collapse of binary neutron star merger event from a cavity of membrane-angular momentum barrier \cite{Abedi:2018npz}. }
\label{echo_pic_2}
\end{figure}

The direct observation of gravitational waves \cite{Abbott:2016blz,TheLIGOScientific:2016src,Abbott:2017vtc,Abbott:2018lct} has provided an unprecedented opportunity to test general relativity (GR) in strong gravity regime. Although, the reported detections were successfully consistent with some predictions of GR \cite{TheLIGOScientific:2016src,Abbott:2017vtc,Abbott:2018lct}, the first tentative search for echoes \cite{Abedi:2016hgu} motivated from resolution of black hole (BH) information problem (in the extreme physical conditions at the BH horizon limit) in the first observing run O1 by Advanced LIGO detectors turns out to be $99\%$ (at false alarm rate probability of $1\%$ or significance of $2.5\sigma$) consistent with Planckian deviations from GR. If correct, this implies that the black hole horizon is not totally absorbing, allowing for postmerger repeating gravitational wave echoes (shown in Fig. \ref{echo_pic_1} for BBH and Fig. \ref{echo_pic_2} for BNS merger via neutron star collapse to a black hole) \cite{Cardoso:2016rao,Cardoso:2016oxy,Cardoso:2019rvt} which are produced in the cavity trapping gravitational waves by angular momentum barrier and near horizon membrane/firewall. 

In this analysis, theoretical best-fit waveform for Hanford and Livingston detectors  $M_{H,I}(t)$, $M_{L,I}(t)$ respectively for the BBH events, provided by the  LIGO and Virgo collaborations and observed data strain from the two detectors, $h_{H,I}(t)$ and $h_{L,I}(t)$ respectively, at 4096 Hz and for 32 sec duration was used. Then based on \cite{Cardoso:2016rao,Cardoso:2016oxy}, a phenomenological gravitational wave template for the echoes using five free parameters was built. The free parameters of the model were
\begin{enumerate}
\item $\Delta t_{\rm echo}$: expected time delay between echoes,  assumed to be with $1\sigma$ error range for Planckian structure. 

\item $t_{\rm echo}=(0.99\Delta t_{\rm echo},1.01\Delta t_{\rm echo})$: time of arrival of the first echo, with 1\% uncertainty due to non-linear dynamics near merger. 

\item $t_0$: varying within the range $t_{0} \in (-0.1,0) \overline{\Delta t}_{\rm echo}$ determines truncation for GR waveform with following smooth cut-off function,
\begin{equation}
\Theta_I(t, t_{0})\equiv\frac{1}{2}\left\{1+ \tanh\left[\frac{1}{2} \omega_I(t)(t-t_{\rm merger}-t_{0})\right]  \right\},
\end{equation}
where $t_{\rm{merger}}$ is defined as the time of peak of template and $\omega_I(t)$ is frequency determined from GR waveform as a function of time \cite{TheLIGOScientific:2016src}. Therefore, the truncated model is defined as,
\begin{eqnarray}
{\cal M}_{T,I}^{H/L} (t, t_{0}) \equiv\Theta_I(t, t_{0}) {\cal M}_{I}^{H/L} (t).
\end{eqnarray}
where H/L are Hanford/Livingston respectively and I represents event name.

\item $\gamma$: which varies within $(0.1,0.9)$, defined as damping factor of successive echoes.

\item $A$: which is to have a flat prior is the over-all amplitude of the echo template, with respect to the merger event.

\end{enumerate}

The full template for echoes in terms of these free parameters is:
\begin{eqnarray}
&&M_{TE,I}^{H/L}(t) \equiv  \nonumber\\
&&A\displaystyle\sum_{n=0}^{\infty}(-1)^{n+1}\gamma^{n} {\cal M}_{T,I}^{H/L}(t+t_{\rm merger}-t_{\rm echo}-n\Delta t_{\rm echo},t_{0}). \label{template} \nonumber \\ 
\end{eqnarray}
This template is commonly referred to as the ADA waveform. 

Following our work, several attempts to replicate/extend this finding were made with positive \cite{Abedi:2016hgu, Conklin:2017lwb, Abedi:2018npz,Uchikata:2019frs, Holdom:2019bdv}, mixed \cite{Westerweck:2017hus,Nielsen:2018lkf,Salemi:2019uea}, and negative \cite{Lo:2018sep,Uchikata:2019frs,Tsang:2019zra} results. So far, the searches for echoes have employed four strategies that can be categorized into:
\begin{enumerate}
\item  Time domain \cite{Abedi:2016hgu,Westerweck:2017hus,Uchikata:2019frs},
\item Frequency domain \cite{Abedi:2018npz,Conklin:2017lwb,Holdom:2019bdv} (using resonances as a consequence of repeating property of echoes),
\item Waveform dependent \cite{Abedi:2016hgu,Westerweck:2017hus,Uchikata:2019frs}, and
\item Model-agnostic or coherent \cite{Abedi:2018npz,Conklin:2017lwb,Salemi:2019uea,Holdom:2019bdv}.  
\end{enumerate}

These searches lead to tentative evidence and detection found with different groups \cite{Abedi:2016hgu, Conklin:2017lwb, Westerweck:2017hus, Abedi:2018npz,Salemi:2019uea,Uchikata:2019frs, Holdom:2019bdv} at false alarm rates of $0.002\%-5\%$ (but see \cite{Westerweck:2017hus, Ashton:2016xff, Abedi:2017isz, Abedi:2018pst,Salemi:2019uea} for the ongoing discussion, comments, and rebuttals on statistical significance of these findings that motivate further investigations). 

Most significant claim was reported by us in \cite{Abedi:2018npz}, where we examined the existence of echoes by building an optimal model-agnostic search strategy via cross-correlating the two detectors in frequency/time in first binary neutron star (BNS) merger event GW170817 in O2.
If a BNS merger event can collapse into a black hole (Fig. \ref{echo_pic_2}), it also provides an opportunity to test GR at the extreme physical condition of the formation of a horizon.  While the current LIGO/Virgo detectors are limited to low frequencies and thus unable to detect BNS classical postmerger signal at $\gtrsim$ kHZ , the echo chamber cavity suppress echoes frequency by a $\ln\left(M/M_{\rm planck}\right) \sim 90$ factor, allowing them to show up squarely within the LIGO sensitivity band.

If confirmed, this detection will be pinpoint both the physics of quantum black holes, and astrophysics of  binary neutron star mergers.

In this paper, we provide a brief status update for the {\textit{Echoes from the Abyss}} \cite{Abedi:2016hgu,Abedi:2018npz} for both BBH and BNS events along with a short review of other positive  \cite{Abedi:2016hgu,Conklin:2017lwb,Abedi:2018npz, Uchikata:2019frs, Holdom:2019bdv}, mixed  \cite{Westerweck:2017hus,Nielsen:2018lkf,Salemi:2019uea}, and negative \cite{Uchikata:2019frs,Tsang:2019zra,Lo:2018sep} search results.

\section{Echoes from the Abyss: Status update O1 \label{Echoes from the Abyss: Status update O1}}
In this section, we bring together all the searches for signals in post-merger gravitational wave data of first observing run O1. We point out their interesting findings and comment on potential misinterpretations. This also helps us paint a possible unified phenomenological picture for echoes. 
\subsection{\label{Five independent groups, Five independent methods, identical results}Five independent groups, Five independent methods, Identical results!}

Searches for post-merger signals consistent with echo predictions using the public data release by the Advanced LIGO gravitational wave observatory has lead to several findings reported by different groups. Here, we highlight the similarities of their finding. However, we caution that these similarities do not guarantee that the signals found are the same (or real), but rather provides support for further investigation.

\begin{enumerate}
\item The time delays of \textbf{0.1 sec} and \textbf{0.2 sec} for post-merger signals for GW151226 and GW151012 respectively found by Salemi et al.  \cite{Salemi:2019uea}, are consistent with time delays first reported in Table II of \cite{Abedi:2016hgu}.

\item Results of \cite{Abedi:2016hgu,Uchikata:2019frs,Westerweck:2017hus,Nielsen:2018lkf,Salemi:2019uea,Abedi:2018npz} show consistency for \textbf{Planckian echoes}, at p-values of $0.002\%-5\%$ for O1 {\it and} O2 events.

\item The reconstructed detector response for post-merger signals of the events GW151226 and GW151012 \cite{SalemiGW151012, SalemiGW151226} in \cite{Salemi:2019uea} show consistent amplitudes \textbf{(0.33, 0.34)$\times$(maximum amplitude of main event)} with what was obtained in Table II  of \textit{Echoes from the Abyss} \cite{Abedi:2016hgu}.

Additionally, energies reported in \cite{Abedi:2016hgu} (Appendix A) is also consistent with strength of signals found by Salemi et al. \cite{Salemi:2019uea}.

Finally, having highest value SNR reported for GW151012 in \cite{Abedi:2016hgu} (Table II and Fig. 6) is also consistent with highest significant event in \cite{Salemi:2019uea}.

\item  Log-Bayes factors for echoes, reported in Table II of Nielsen et al. \cite{Nielsen:2018lkf}  for GW151012 and GW151226 having GW151012 as highest significant is also consistent with ordering of significance of signals found by Salemi et al. \cite{Salemi:2019uea}.

\item Noting that echo signal of GW150914 \cite{Abedi:2016hgu} at time delay 0.3 sec had narrowest time window ($\pm 3\%$ in table II) and smallest energy (Table II in \cite{Abedi:2016hgu} ) compared to GW151226 and GW151012 is consistent with non-detection by Salemi et al. \cite{Salemi:2019uea} and the negative Log Base factor by Nielsen et al.  (Table II in \cite{Nielsen:2018lkf}). 

\item Interestingly, the residual signal for GW150914 \cite{SalemiGW150914} in supporting material of Salemi et al. \cite{Salemi:2019uea} is consistent with \textbf{300 msec} echo signal time delay reported in Table II of \textit{Echoes from the Abyss} \cite{Abedi:2016hgu}. The fact that the postmerger signal only appears at a single pixel, suggests the weakness of the signal, as mentioned above. 

\item Furthermore, GW151012 as the most significant echoes signal obtained by Westerweck et al. \cite{Westerweck:2017hus} (p-value of 6\% in their Table I) is also consistent with the most significant signal found by Salemi et al. \cite{Salemi:2019uea} (p-value of 0.4\%).

\item The echo search results of Uchikata et al. \cite{Uchikata:2019frs} (using ADA waveform in Appendix A) for O1 events shown in Table \ref{table_3} are consistent with results from other groups  \cite{Abedi:2016hgu,Westerweck:2017hus,Nielsen:2018lkf,Salemi:2019uea}.


\item Table IV of Lo et al. \cite{Lo:2018sep}, which adds the GR waveform for the main event to the ADA waveform, keeping only three echoes with larger prior ranges also found similar ordering of events by their statistical significance and p-values, comparable to what was reported by Nielsen et al.  \cite{Nielsen:2018lkf} and Westerweck et al. \cite{Westerweck:2017hus}.

\end{enumerate}

\subsection{\label{Hint}Evidence for dependence of significance of echoes on binary BH mass ratio,\\ \textit{Comment on: Salemi et al.} \cite{Salemi:2019uea}}

In this part, we comment on the conclusion of Salemi et al. \cite{Salemi:2019uea} about the post-merger signal found in GW151012, which they claim to have arrived from a different sky location than the main event. We then provide possible evidence for how mass ratio of LIGO BBH events are correlated with the significance of the echo signals they report.

\subsubsection{Comment on: Salemi et al. \cite{Salemi:2019uea}}

Salemi et al. \cite{Salemi:2019uea} claimed that the significant postmerger signal they see for GW151012 has arrived from a different sky location. Although posterior probability of time delay between Hanford and Livingston (Fig. 4 of \cite{Salemi:2019uea}) is consistent with this claim (which is a posteriori statistics), it would be very unlikely from their reported post-merger signal statistical significance p-value$\simeq 0.004$ that assumes spatial coincident with the main event. We also note that all the secondary (post-merger) clusters they claim (and search for) as signals in their Figs. 3 and 5 for GW151012 and GW151226 respectively are monochromatic. This would clearly lead to a degeneracy in the inference of time-delay. We note that in the secondary signal of GW151012, the null (residual) plot of Salemi et al. \cite{Salemi:2019uea} in Fig. \ref{Salemi4} appears to have the dominant peak of the cluster (mainly causing a different sky localization) at $\sim 130$ Hz corresponding to a 7.7 msec period. Interestingly, as illustrated in Fig. \ref{Salemi4}, this 7.7 msec is the same as the time-delay between the first peak (on-source) and second peak (off-source) for the post-merger signal (green). This implies that, as we see in Fig. \ref{Salemi4}, this monochromatic degeneracy might be responsible for the wrong sky localization.

\textbf{\begin{figure}[!tbp]
\centering
    \includegraphics[width=0.4\textwidth]{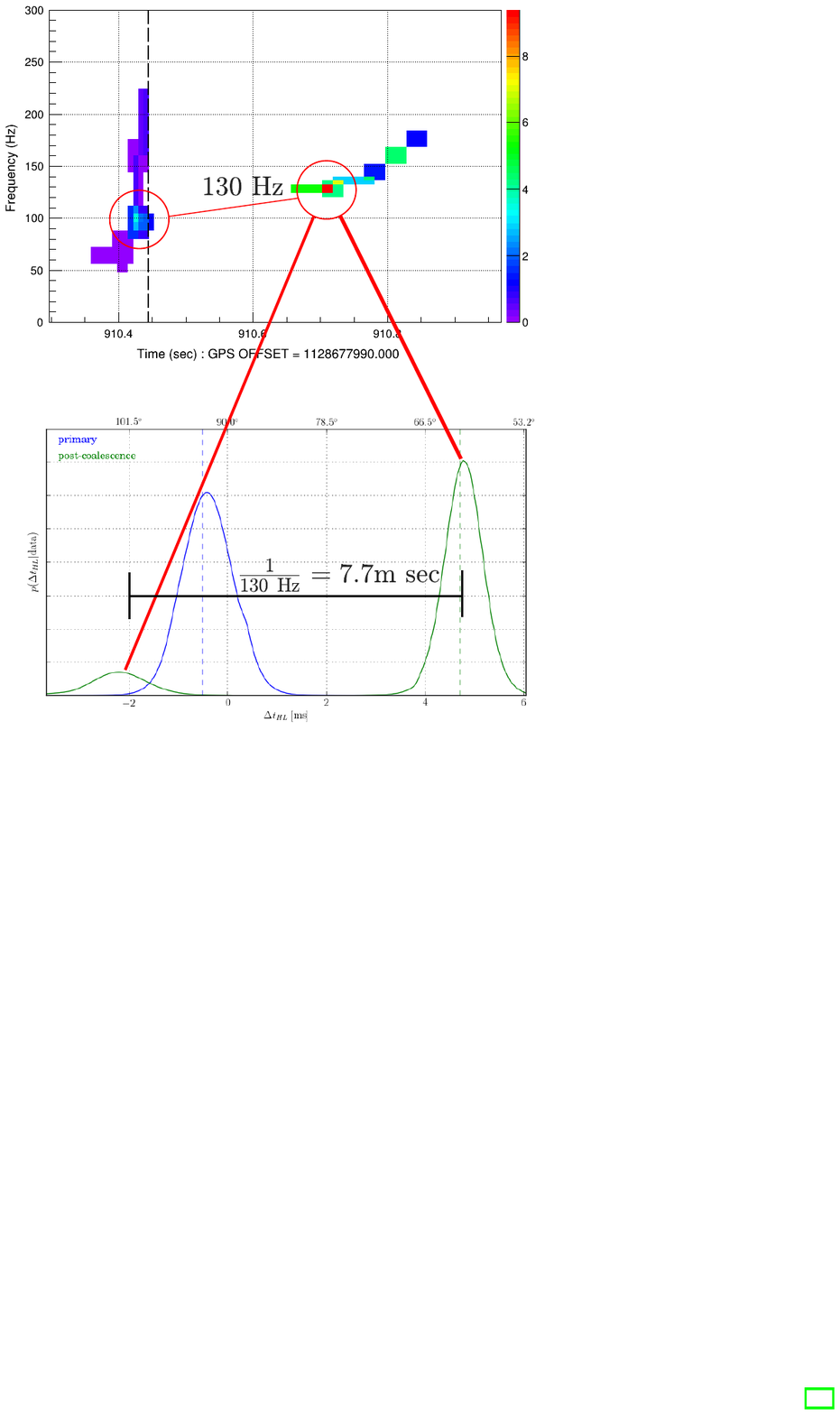}
 \caption{(Residual) cWB waveform reconstruction of GW151012 and maximum posterior time delay for the main event (blue) and post-coalescence (Echo) signal (green) found by Salemi et al. \cite{Salemi:2019uea}. In these plots (top: residual (null) \& bottom: maximum posterior time delay) additional clues show how wavelength degeneracy can cause 7.7 msec $=1/(130\ \rm Hz)$ shift in maximum posteriori probability for time delay. 
 \label{Salemi4}}
\end{figure}}

%

\subsubsection{Evidence for dependence of significance of echoes on binary BH mass ratio}

The initial of conditions of binary BH mergers, and thus the relative amplitude of echoes may depend on the BBH mass ratio. Here, we suggest that there is evidence for this dependence in the results reported by Salemi et al. \cite{Salemi:2019uea}.

\begin{enumerate}
    \item \textit{Mass ratio versus p-value}:
    
   In the following, we outline the method that leads to our finding of mass ratio dependence of significance of echoes (using p-values reported by Salemi et al. \cite{Salemi:2019uea}),
    \begin{enumerate}
        \item We determine mass ratios of BBH events using LIGO parameter estimation samples for \cite{GWTC-1} and weighting all events as equal. A full $m_{1}$ vs $m_{2}$ distribution samples for ``Overall\_posterior'' is used. The blue errorbars in Fig. \ref{mass ratio error} show the 50\% confidence regions for $m_2/m_1$ mass ratio for the 10 BBH events in O1 and O2 (Table \ref{table_20}).

        \item Best-fit straight line of mass ratio vs $\sqrt{-\log(p-value)}$ is plotted taking p-values reported by Salemi et al. \cite{Salemi:2019uea} for postmerger signal of each event. Here, least square method \cite{least-square-method} is used to fit a straight line. We use the slope of the best-fit line as our measure of correlation, taking into account all the posterior points of mass ratios for all events.

        \item Accordingly, the significance of correlation in Fig. \ref{histogram-slope} is obtained, compared to the null hypothesis that there is no relation between p-value and mass-ratio by randomly assigning p-values for BBH Catalog events within the uniform range $0<\sqrt{-\log(\rm{p-value})}<2.5$ ($1>$ p-values$ >0.0019$). Since, no relation between p-value and mass ratio of events must lead to zero mean slope in large number of random selections, we estimate the significance of correlation using the fraction of randomized slopes that are higher than the actual measurement.  Fig. \ref{histogram-slope} shows that we find tentative evidence of mass-ratio dependence of p-values reported by Salemi et al. \cite{Salemi:2019uea} at at false detection probability of 1\%.

    \end{enumerate}
    
    \begin{figure}
    \centering
    \includegraphics[width=0.4\textwidth]{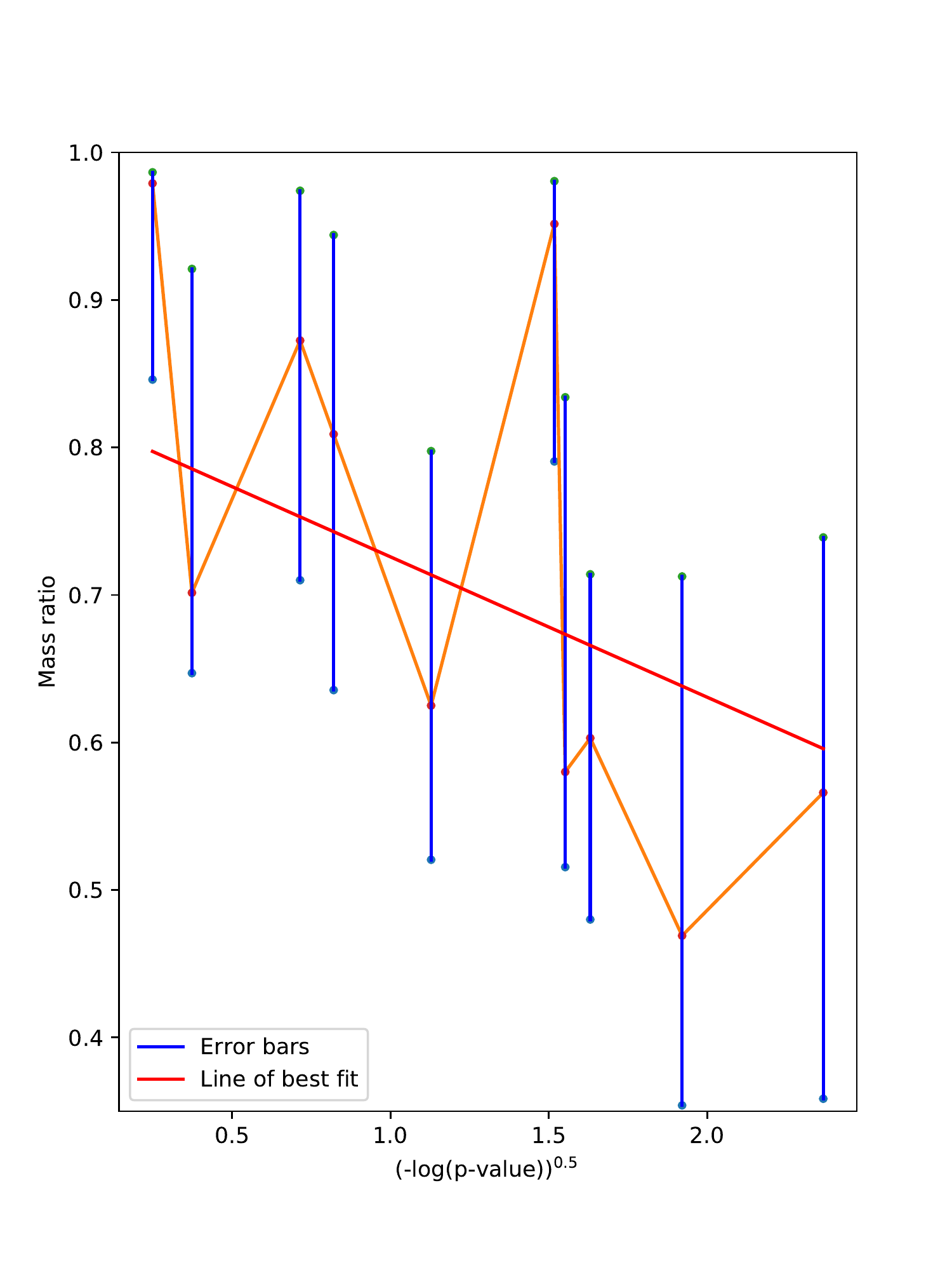}
    \caption{Plot of mass ratio vs $\sim \sqrt{-\log(\rm p-value)}$ (where p-values reported by Salemi et al. \cite{Salemi:2019uea}). Vertical lines indicating error bars for 50\% credible region and central points are most likelyhood value of mass ratio given from posterior distribution. The relation of p-value to error function erf(SNR) requires to take roughly SNR $\sim \sqrt{-\log(\rm p-value)}$ as horizontal axis. The ``line of best fit'' indicates the best fit to using all the mass ratio posterior points, using the same weighting for all events.}
    \label{mass ratio error}
    \end{figure}

    \begin{figure}
    \centering
    \includegraphics[width=0.4\textwidth]{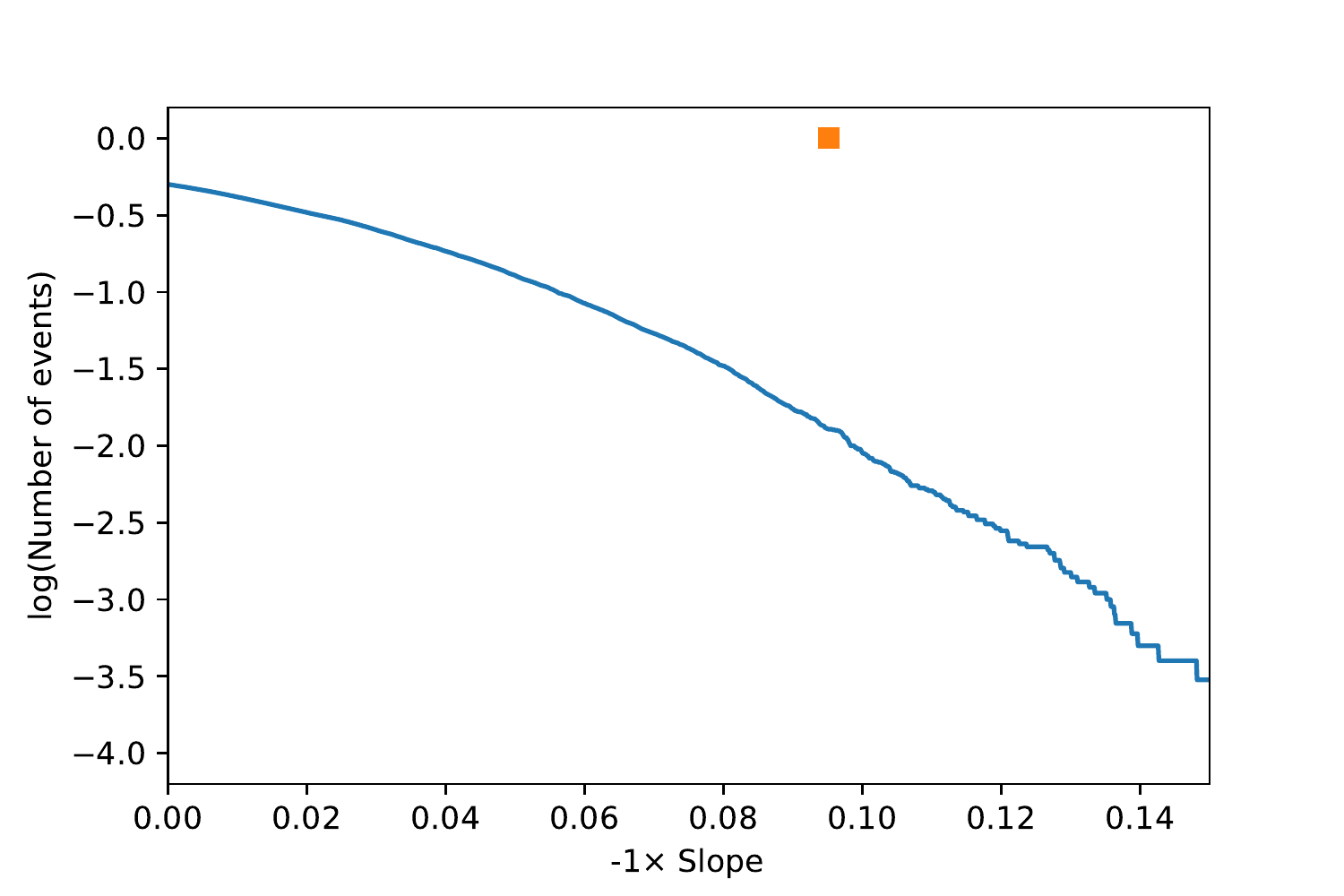}
    \caption{Plot of histogram of slopes considering uniform random selection of $0<\sqrt{-\log(\rm{p-value})}<2.5$. The histogram shows the false alarm rate is 0.0128.}
    \label{histogram-slope}
    \end{figure}
    
    

\end{enumerate}

Table \ref{table_20} indicates events and p-values reported by Salemi et al. \cite{Salemi:2019uea} versus average mass ratios. It appears that smallest mass ratios correspond to smallest p-values. As an alternate rough estimate, we have 1/10 chance out of 10 BBH events that smallest mass ratio goes to smallest p-value. Removing this event, we get 1/9 chance for the same occurrence at random. Therefore, for the two most significant events in Table \ref{table_20}, the chance of getting the most extreme mass ratio becomes p-value$=\frac{1}{9}\times \frac{1}{10} = 0.011$ which is also consistent with results found in Fig. \ref{histogram-slope}. 

While the latter, on its own, would have been an {\it a posteriori} analysis (and thus not conclusive), it supports the earlier results based on the distribution of slopes, and thus adds further weight to the conclusion that postmerger signals found by Salemi et al. \cite{Salemi:2019uea} are correlated with the mass ratio of the BBH mergers. 

\begin{table}[!ht]
\begin{center}
\begin{tabular}{ |c|c|c|c|c| }
\hline
Event & p-value $\pm 2\sigma$ & average mass ratio  \\
\hline
GW150914 & $0.94 \pm 0.02$& 0.86  \\
\cellcolor{blue!25} GW151012& \cellcolor{blue!25} $0.0037 \pm 0.0014$& \cellcolor{blue!25} 0.58 \\
\cellcolor{blue!25} GW151226& \cellcolor{blue!25}$0.025 \pm 0.005$ & \cellcolor{blue!25}0.56 \\
GW170104& $0.07 \pm 0.01$& 0.65 \\
GW170608& $0.51 \pm 0.02$& 0.68 \\
GW170729& $0.09 \pm 0.01$& 0.68 \\
GW170809& $0.28 \pm 0.01$& 0.68\\
GW170814& $0.10 \pm 0.01$& 0.82\\
GW170818& $0.87 \pm 0.02$& 0.75\\
GW170823& $0.60 \pm 0.02$& 0.74\\

\hline
\end{tabular}
\caption{Events and p-values versus average mass ratios for post-coalescence signals reported by Salemi et al. \cite{Salemi:2019uea}. The highlighted rows indicate the most significant postmerger signals reported in  \cite{Salemi:2019uea}. We see that the most significant signals come from most extreme mass ratios.}\label{table_20}
\end{center}
\begin{center}
\end{center}
\end{table}

\section{\label{Echoes from the Abyss: Status update on LIGO/Virgo O2 and independent astrophysical considerations}Echoes from the Abyss: Status update O2 and independent astrophysical considerations}
If a statistical detection corresponds to a real physical effect, then it should be replicated in independent observations of similar physical phenomena. Furthermore, the conclusions are objective, only if they can be reproduced by independent groups, using independent methods.  Here, we give a brief status update on independent searches for ADA  waveform \cite{Abedi:2016hgu} in O2 BBH events, and postmerger BH formation signal \cite{Abedi:2018npz} in the BNS event (GW170817 in O2) from astrophysical considerations.
\subsection{Uchikata et al. \cite{Uchikata:2019frs} search for ADA waveform \cite{Abedi:2016hgu} in O1 and O2}

Uchikata et al. \cite{Uchikata:2019frs} (Appendix A) have also searched for the ADA waveform (Equation \ref{template}) \cite{Abedi:2016hgu} except that they set the cut-off parameter $t_{0}$ (described in Introduction, above) as fixed and  set the search region of $\Delta t_{echo}$ to the 90\% credible regions (Table I in \cite{Uchikata:2019frs}) of $(a, M)$ for the nine BBH gravitational wave events in O1 and O2 observed by Advanced LIGO and Virgo. They first look for echoes in O1 \href{https://arxiv.org/abs/1906.00838v1}{arXiv:1906.00838v1}, then they extend the analysis to O2. They point out that fixing $t_{0} = -0.1\Delta t_{echo}$ weakly affects SNR and has advantage in saving computational costs. Additionally, $t_{0} = -0.1\Delta t_{echo}$ is the best fit value of O1 echoes. They also set a critical p-value, where below (above) p-value=0.05, echo signals are likely (unlikely) to be present in the data. It is worth noting that their attempt with their own waveform (which cuts off the low-frequency part of ADA waveform) finds no evidence for echoes.

Their results are explained below:

\begin{enumerate}
\item \textit{O1 events (reanalysis of Westerweck et al.} \cite{Westerweck:2017hus}):

Uchikata et al. \cite{Uchikata:2019frs} reproduce the same background estimation as Westerweck et al. \cite{Westerweck:2017hus}. Therefore their p-value results and Poisson errors are compared to their O1 results shown in Table \ref{table_3}. 
\begin{table}[!ht]
\begin{center}
\begin{tabular}{ |c|c|c| }
\hline
Event & Westerweck et al. \cite{Westerweck:2017hus} & Uchikata et al. \cite{Uchikata:2019frs}  \\
\hline
GW150914 & $0.238 \pm 0.043$ & $0.157 \pm 0.035$ \\
\hline
GW151012 & $0.063 \pm 0.022$ & $0.047 \pm 0.019$ \\
\hline
GW151226 & $0.476 \pm 0.061$ & $0.598 \pm 0.069$ \\
\hline
Total & $0.032 \pm 0.016$ & $0.055 \pm 0.021$ \\
\hline
\end{tabular}
\caption{P-values along with Poisson errors for ADA waveform searches in O1 events \cite{Uchikata:2019frs}. The results are consistent within the Poisson errors for all events.}\label{table_3}
\end{center}
\begin{center}
\end{center}
\end{table}

\item \textit{O2 events}:

Uchikata et al. \cite{Uchikata:2019frs} then examined O2 events using ADA waveform \cite{Abedi:2016hgu}. Results of their p-values are presented in Table \ref{table_4}  showing similarly small values, comparable to O1. The total p-value for the six O2 events is 0.039.  This is very significant, since the ADA waveform was first developed and used on O1 data, and thus the small p-value using O2 data, which is completely independent, severely reduces chances of this being a statistical fluke. 

The combined O2 with O1 events gives total p-value=0.047. 

\begin{table}[!ht]
\begin{center}
\begin{tabular}{ |c|c|c| }
\hline
Event & Uchikata et al. \cite{Uchikata:2019frs}  \\
\hline
GW170104 & 0.071 \\
\hline
GW170608 & 0.079 \\
\hline
GW170729 & 0.567 \\
\hline 
GW170814 & 0.024 \\
\hline
GW170818 & 0.929 \\
\hline
GW170823 & 0.055 \\
\hline
Total & 0.039 \\
\hline
\end{tabular}
\caption{P-values for O2 events \cite{Uchikata:2019frs}. The results show O2 events have same small p-values as O1.}\label{table_4}
\end{center}
\begin{center}
\end{center}
\end{table}

\end{enumerate}

\subsection{Status update on the GW170817 postmerger echoes: Electromagnetic confirmation}

\begin{figure}[!tbp]
\centering
    \includegraphics[width=0.4\textwidth]{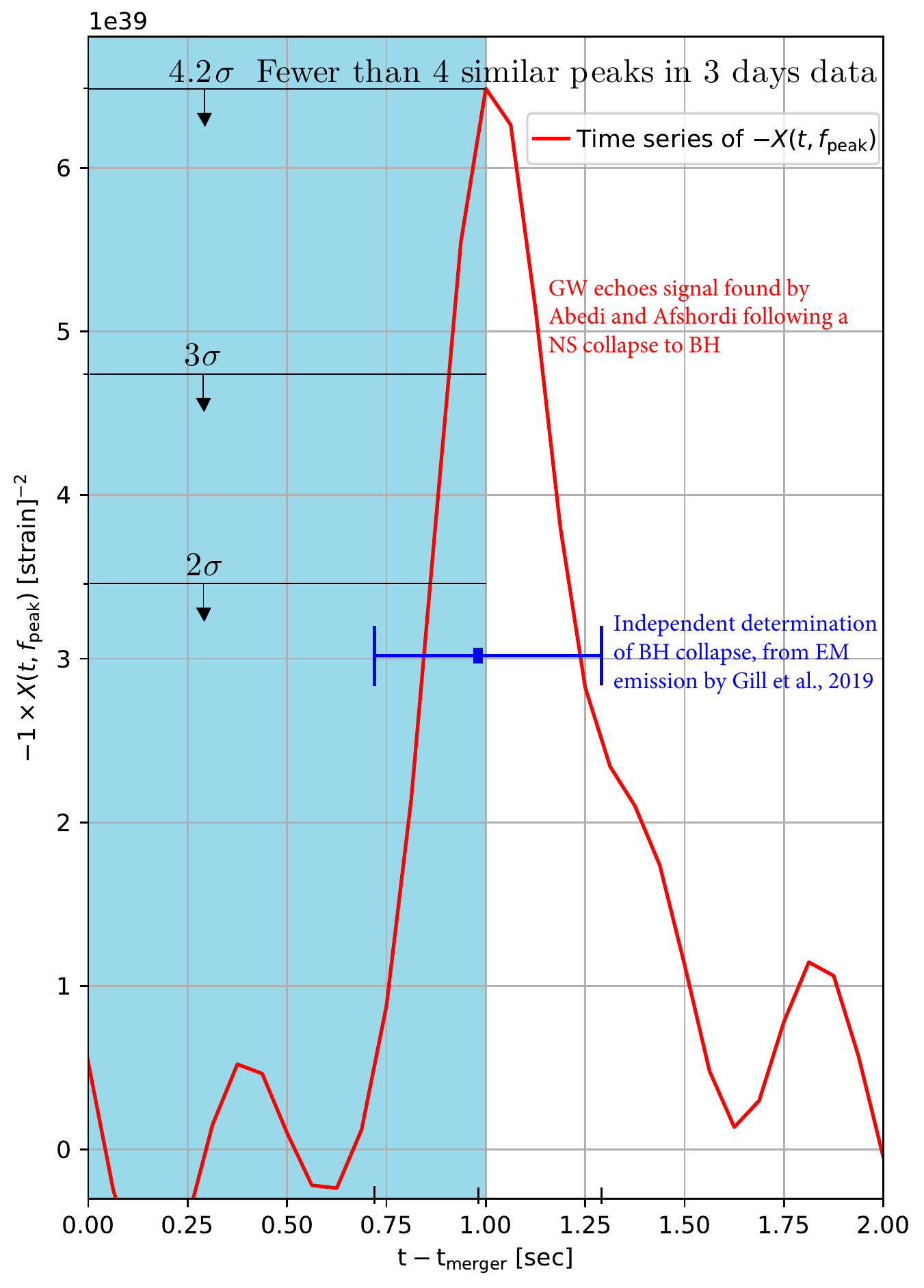}
 \caption{Amplitude-time plot of echo signal found at 1.0 sec and frequency of 72 Hz \cite{Abedi:2018npz} after the merger. Gill et al. \cite{Gill:2019bvq} use independent Astrophysical considerations to argue that the remnant of GW170817 must have collapsed into a BH at $t_{\rm coll}=0.98^{+0.31}_{-0.26}$ sec. The same timescale for collapse was reported first by Abedi and Afshordi \cite{Abedi:2018npz}. Error-bar (in blue) is the time of collapse considering this independent observation in \cite{Gill:2019bvq} compared to the detected signal of echoes which is also expected from formation quantum BHs. The two independent observations happen to coincide perfectly! (The shaded region is 0-1 sec prior range after the merger, used for p-value estimate of $1.6 \times 10^{-5}$.)}
 \label{NS-NS_9}
\end{figure}

In \cite{Abedi:2018npz}, we used a model-agnostic search to find tentative detection of echoes at $f_{echo}\simeq 72$ Hz, around 1.0 sec after the BNS merger event GW170817. Gill et al. \cite{Gill:2019bvq}, using completely independent Astrophysical considerations, have also determined that the remnant of GW170817 must have collapsed into a BH at $t_{\rm coll}=0.98^{+0.31}_{-0.26}$ sec. This timescale for collapse was first reported by Abedi and Afshordi in \cite{Abedi:2018npz}. Error-bar for this observation along with the detected GW signal of echo as a consequence of BH collapse is shown in Fig. \ref{NS-NS_9} as comparison. We see that these independent observations (one based on GW echoes, and another inferred from electromagnetic signals) happen to coincide. 

The false alarm rate for this GW echo signal is p-value=$1.6\times10^{-5}$ considering all the "look-elsewhere" effects \cite{Abedi:2018npz}. In other words, a similar postmerger signal inside the anticipated frequency/time window for echoes cannot occur more than 4 times in 3 days from detector noise, and yet it happens within 1 second of the BNS merger. Its coincidence with the inferred BH collapse time from electromagnetic observations provides \cite{Gill:2019bvq} strong and independent evidence for this detection. 

\section{\label{Conclusion}Conclusion and Discussions}

A scientific and conclusive search for echoes should (arguably) satisfy the following three criteria:

\begin{enumerate}

\item be based on a proper physical, or physically-motivated, model (otherwise, it might miss the signal)

\item be simple (avoids using too many arbitrary choices, and/or free parameters)

\item avoids {\it a posteriori} statistics (a model built to a fit a dataset, would of course fit that dataset!)

\end{enumerate}

The scope of this paper was to give an update on the first search for echoes by Abedi, Dykaar, and Afshordi \cite{Abedi:2016hgu}, on its 3rd anniversary. 

Section \ref{Echoes from the Abyss: Status update O1} argues that a unified picture does emerge from the analyses by independent groups, and demonstrates that they all find consistent evidence for ADA echoes, at p-values of few percent. We also provide statistical evidence for why echoes might be more prominent for extreme mass-ratio BBH events. This is why the amplitudes of echoes might be lower in O2, compared to O1. This observation may provide a new insight into proper physical modelling of echoes, using e.g., Effective One Body (EOB) formalism. 
 
Section \ref{Echoes from the Abyss: Status update on LIGO/Virgo O2 and independent astrophysical considerations} provides a status update of echo searches in the second LIGO/Virgo observing run (O2) by independent groups. In summary, it appears that independent searches do provide significant evidence for echoes in the O2 BBH events (\cite{Uchikata:2019frs} consistent with  \cite{Abedi:2016hgu}, despite using independent data) and the BNS event (\cite{Gill:2019bvq} consistent with \cite{Abedi:2018npz}, despite using independent probes). 

Let us close by some technical points for {\it echology aficionados} \cite{Wang:2018gin}:

\begin{enumerate}

\item{\textit{Binary black hole mergers:}}
Uchikata et al. \cite{Uchikata:2019frs} used ADA waveform \ref{template} \cite{Abedi:2016hgu} with $t_{0}=-0.1 \Delta t_{echo}$ fixed, with best fit value of O1 and having search in 90\% credible region (Table I in \cite{Uchikata:2019frs}) of $(a,M)$ to set $\Delta t_{echo}$, looked for echoes for both O1 and O2. The results for O2 (with p-value=0.039) presented in Table \ref{table_4} indicating similar evidence as O1 (with p-value=0.055) Table \ref{table_3}. Using their search strategy and prior, but our original proposed background estimation \cite{Abedi:2016hgu} we can confirm that same evidence in combined O1 and O2 events. We also confirm that fixing $t_{0}=-0.1 \Delta t_{echo}$ does not affect the result which is what they use to reduce the computational costs. Moreover, we shall point out that fixing $\gamma=0.9$ also does not affect the result as these parameters are not treated as independent in ADA search  and were treated as universal in combining several events.

\item{\textit{Binary neutron star merger:}}
Gill et al. \cite{Gill:2019bvq} with independent astrophysical consideration have determined that the remnant of GW170817 must have collapsed to a BH after $t_{\rm coll}=0.98^{+0.31}_{-0.26}$ sec. This time is consistent with what we already reported in \cite{Abedi:2018npz} as collapse to black hole via detection of echo signal with $4.2\sigma$ significance. The echo signal and error-bar for this observation are compared in Fig. \ref{NS-NS_9}. The question that remains is whether astrophysical black holes with significant accretion (such as GW170817 remnant) should have similar echo properties as the quantum black holes in vacuum. Another possibility (already discussed in \cite{Abedi:2018npz}) is that the detected ``echo'' signal is due to extremely narrow quasi-periodic oscillations in the BH accretion disk (and not a signature of quantum BH horizons).

\item \textit{Concerns about errors in $\Delta t_{echo}$}: The original search of ADA \cite{Abedi:2016hgu} determined errors for $\Delta t_{echo}$ using an ad-hoc method (as the posterior distributions was not public on the time of their research) giving symmetric 1-sigma errors. Although, we have public posterior distributions, still we do not know how much error comes from instrumental and how much from systematics.  Additionally, this method (1-sigma errors) misses one third of the signals. Since loud events would have shrinking systematic errors, one must modify the method taking into account the uncertainty in the scale of quantum gravity (e.g., Planck length, versus reduced Planck length). However, too wide a prior can bury the signal in the noise. Therefore, a search strategy based on a more physical model (e.g., \cite{Wang:2019rcf}), using proper Bayesian methodology can improve the efficacy of the current searches. 


\item \textit{Concerns about keeping $t_{0}$ and $\gamma=0.9$ fixed}: Uchikata et al. \cite{Uchikata:2019frs} have searched in O2 having $t_{0}$ fixed at its best fit value in O1. This keeping the parameters in their best fit value might be a good idea to make speed up the search, but can miss echoes in significant event, where spin and mass errors are small. 

\item \textit{Concerns about mass ratio dependence of echoes overall amplitude}: In Section \ref{Hint}, we found that significance of echoes appear to depend on the BBH progenitor mass ratio (Fig. \ref{mass ratio error}). However, it is hard to tell whether this could be a physical effect, or an artifact of the cWB search strategy employed by Salemi et al. \cite{Salemi:2019uea}. 

\end{enumerate}

Finally, we again point out that an optimal search must use simplest model having minimum free parameters. Therefore, it would be reasonable that Tsang et al. \cite{Tsang:2018uie,Tsang:2019zra} did not find evidence for echoes, as their nominal model composed of 5 sine-Gaussians has 49 free parameters, and requires SNR$>$8 for detection.  In contrast, all the searches that find evidence for echoes used $\leq$ 5 free parameters, and recover SNR$\sim 4$. For the same reason, it would be unlikely to obtain any signal by significantly increasing prior ranges, although it may covers more possibilities. This would explain lower significance reported by Lo et al. \cite{Lo:2018sep}.  Additionally, unlike the original ADA waveform  \ref{template} \cite{Abedi:2016hgu}, the failed search of Uchikata et al. \cite{Uchikata:2019frs} using their proposed model cuts off low frequencies which appear to be crucial in recovering the echo signal (as independently found in  \cite{Abedi:2018npz}). Indeed, this is also explicitly recognized by \cite{Uchikata:2019frs}. 

We conclude by stating the obvious: With more theory, data and statistical methodologies on the way, the search for black hole gravitational wave echoes remains extremely confusing, active and exciting. So, stay tuned (or join in)!

\section{Acknowledgement}
JA would like to thank Bruce Allen, Badri Krishnan, and Alex Nielsen for all of their supports. We thank the Max Planck Gesellschaft and the Atlas cluster computing team at AEI Hannover for support and computational help.
This work was partially supported by the University of Waterloo, Natural Sciences and Engineering Research Council of Canada (NSERC), the Perimeter Institute for Theoretical Physics. Research at the Perimeter Institute is supported by the Government of Canada through Industry Canada, and by the Province of Ontario through the Ministry of Research and Innovation.
This research has made use of data, software and/or web tools obtained from the Gravitational Wave Open Science Center (https://www.gw- openscience.org), a service of LIGO Laboratory, the LIGO Scientific Collaboration and the Virgo Collaboration. LIGO is funded by the U.S. National Science Foundation. Virgo is funded by the French Centre National de Recherche Scientifique (CNRS), the Italian Instituto Nazionale della Fisica Nucleare (INFN) and the Dutch Nikhef, with contributions by Polish and Hungarian institutes.

\bibliography{Paper}
\end{document}